\documentclass[sigconf]{acmart}

\usepackage[backend=bibtex,style=verbose-trad2,style=numeric,
citestyle=numeric]{biblatex}
\AtEveryBibitem{%
  \clearfield{issn} 
  \clearfield{doi} 
  \clearfield{isbn} 
  \clearfield{eprint} 
  \clearfield{publisher} 
  \clearfield{location}
  \clearfield{editor}
  \clearfield{address}
  \clearfield{volume}
  \clearfield{number}
  \ifentrytype{online}{}{
    \clearfield{url}
  }
}
\usepackage{amsmath,amsfonts}
\usepackage{algorithmic}
\usepackage{graphicx}
\usepackage{textcomp}
\usepackage{xcolor}
\usepackage{listings}
\def\BibTeX{{\rm B\kern-.05em{\sc i\kern-.025em b}\kern-.08em
    T\kern-.1667em\lower.7ex\hbox{E}\kern-.125emX}}
\usepackage{sansmath}
\usepackage{hyperref}
\usepackage{tikz}
\usepackage{tcolorbox}
\usepackage{nicefrac}
\usepackage{balance}

\usepackage{amsthm}
\usepackage{lipsum}
\usepackage{subcaption}
\usepackage{float}
\usepackage{booktabs}
\usepackage{xspace}
\usepackage{pgfplots}
\usepackage{tikz}
\usepackage{adjustbox}
\usepackage{subcaption}
\usepackage{cleveref}
\usepackage{wrapfig}
\usepackage{physics}
\usepackage{mwe}

\usepackage{caption} 
\usepackage[vlined, boxruled, linesnumbered] {algorithm2e}
\SetKwComment{Comment}{$\triangleright$\ }{}
\SetCommentSty{itshape}
\newboolean{showcomments}
\setboolean{showcomments}{true}
\ifthenelse{\boolean{showcomments}}
{
	\definecolor{myyellow}{RGB}{255, 228, 26}
    \definecolor{mygreen}{RGB}{173,255,47}
	\definecolor{myblue}{RGB}{50, 50, 220}
	\newcommand{\nb}[2]{
		{\sf
			\fcolorbox{myyellow}{yellow}{\scriptsize\textbf{#1}}%
			$\blacktriangleright$%
			{\color{myblue}\fontsize{8pt}{8pt}\selectfont\textbf{#2}}%
		}%
	}
 	\newcommand{\nk}[2]{
		{\sf
			\fcolorbox{mygreen}{mygreen}{\scriptsize\textbf{#1}}%
			$\blacktriangleright$%
			{\color{myblue}\fontsize{8pt}{8pt}\selectfont\textbf{#2}}%
		}%
	}
}
{
	\newcommand{\nb}[2]{}
    \newcommand{\nk}[2]{}
}

\newcommand{\nsga}{NSGA-II\xspace} %
\newcommand{\nsgadt}{NSGA-II-DT\xspace} %
\newcommand{\rs}{random search\xspace} %

\usepackage[vlined, boxruled, linesnumbered] {algorithm2e}
\SetKw{KwBy}{by}
\SetKw{KwBreak}{break}
\SetKw{KwReturn}{return}
\def\HiLi{\leavevmode\rlap{\hbox to \hsize{\color{gray!35}\leaders\hrule height .8\baselineskip depth .5ex\hfill}}}

\bibliography{papers}

\newcommand{\algo}{\textsc{NSGA-II-SVM}\xspace}
\newcommand{\algolong}{Non-dominated Sorting Genetic Algorithm with Support Vector Machine Guidance}

\usepackage{soul}

\definecolor{codegreen}{rgb}{0,0.6,0}
\definecolor{codegray}{rgb}{0.5,0.5,0.5}
\definecolor{codepurple}{rgb}{0.58,0,0.82}
\definecolor{backcolour}{rgb}{0.95,0.95,0.92}

\lstdefinestyle{mystyle}{
    backgroundcolor=\color{backcolour},   
    commentstyle=\color{codegreen},
    keywordstyle=\color{magenta},
    numberstyle=\tiny\color{codegray},
    stringstyle=\color{codepurple},
    basicstyle=\ttfamily\footnotesize,
    breakatwhitespace=false,         
    breaklines=true,                 
    captionpos=b,                    
    keepspaces=false,                 
    numbers=left,                    
    numbersep=3pt,                  
    showspaces=false,                
    showstringspaces=false,
    showtabs=false,                  
    tabsize=2,
    language=python,
    numbers=left,
    stepnumber=1
}

\newtcolorbox{rq_answer}[1][]{interior hidden,
colback=white,#1}

\usepackage{draftwatermark}
\SetWatermarkLightness{0.95}
\SetWatermarkText{Preprint}
\SetWatermarkScale{1}

\setcopyright{acmlicensed}
\copyrightyear{2024}
\acmYear{2024}
\acmDOI{XXXXXXX.XXXXXXX}

\acmConference[ICSE '24]{46th International Conference on Software Engineering}{April 14--20,
  2024}{Lisbon, Portugal}
  
\begin{document}

\author{Lev Sorokin}
\affiliation{%
  \institution{fortiss, Research Institute of the Free State of Bavaria}
  \streetaddress{Guerickestraße 25, 80805 Munich, Germany}
  \city{Munich}
  \country{Germany}}
\email{sorokin@fortiss.org}

\author{Niklas Kerscher}
\affiliation{%
  \institution{Technische Universität München}
  \city{Munich}
  \country{Germany}}
\email{niklas.kerscher@tum.de}

\pagenumbering{arabic} 
\pagestyle{plain}
\settopmatter{printacmref=True}


\title{Guiding the Search Towards Failure-Inducing Test Inputs Using Support Vector Machines}

\settopmatter{printacmref=false}

\begin{abstract}
\label{sec:abstract}
In this paper, we present \algo (\algolong), a novel learnable evolutionary and search-based testing algorithm that leverages Support Vector Machine (SVM) classification models to direct the search towards failure-revealing test inputs. Supported by genetic search,  \algo creates iteratively SVM-based models of the test input space, learning which regions in the search space are promising to be explored. A subsequent sampling and repetition of evolutionary search iterations allow to refine and make the model more accurate in the prediction. Our preliminary evaluation of \algo by testing an Automated Valet Parking system shows that \algo is more effective in identifying more critical test cases than a state of the art learnable evolutionary testing technique as well as na\"ive \rs.
\vspace*{-0.05cm}
\end{abstract}

\keywords{Search-based Software Engineering, Scenario-based testing, Automated Driving, Machine Learning, Support-vector Machines}

\maketitle


\vspace*{-3pt}\section{Introduction}
\label{sec:intro}
Search-based software testing (SBST) is an effective approach for testing systems 
with large and complex input spaces, such as learning-enabled automated driving or avionic systems \cite{MenghiApproxCPS2020,Moghadam21Deeper,Riccio2020DeepJanus, Zohdinasab21DeepHyperion} to avoid expensive and dangerous real-world testing.
SBST relies on meta-heuristics, such as evolutionary algorithms~\cite{BorgANJS21, humeniuk2022searchbased,Klück19Nsga2ADAS,Riccio2020DeepJanus} which consider the testing problem as a multi-objective optimization (MOO) problem, where system-level safety requirements are modeled by different objectives to be optimized, such as time-to-collision or safety distance. To execute SBST, the system-under-test (SUT) is in general run in a simulation environment \cite{Dosovitskiy17Carla, Prescan} and exposed to different scenarios, defined by behaviors of other actors or environmental conditions. 

However, executing a SUT in a simulator can take minutes to hours, making testing with multiple scenario parameters and large parameter spaces computationally expensive \cite{nejati2023reflections}.
One promising approach to make the search more efficient is combining evolutionary search with surrogate models \cite{Raja16NeuralNSGA2, nejati2023reflections}. Surrogate models can enrich testing by 1) predicting simulation results to avoid executing the SUT in the simulator \cite{Raja16NeuralNSGA2,Haq2022SurrogateMOP}, or 
2) by employing machine-learning (ML) models to predict most likely failure-revealing regions of the test inputs space to evolve the search using those \cite{Raja18NSGA2DT}.

\begin{figure}[!tbp]
    \centering
    
    \includegraphics[width=0.23\textwidth]{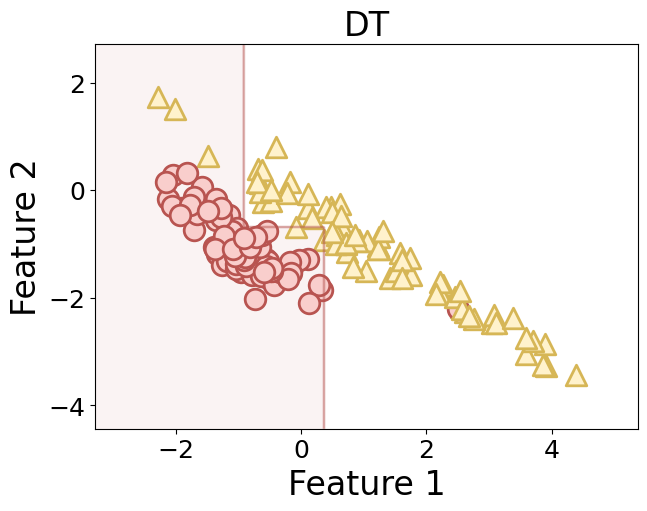}
    \includegraphics[width=0.23\textwidth]{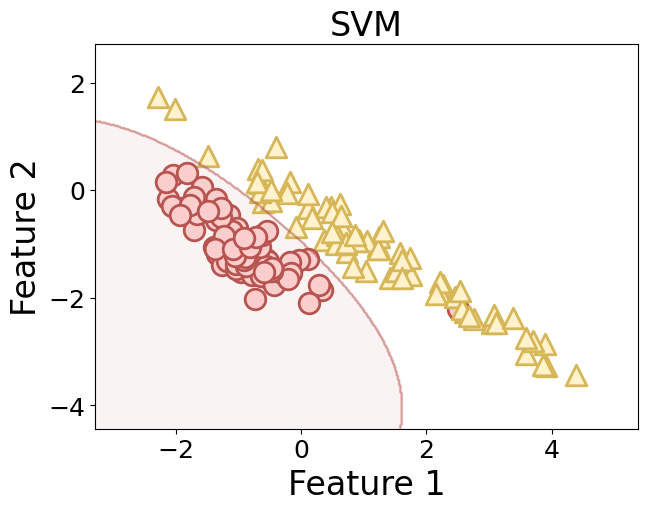}
    
    \caption{Illustration of difference between boundary identification by a Decision Tree and Support Vector Machine for data from two classes (circle/triangle positive/negative data).}
    \label{fig:motivation}
\end{figure}

Abdessalam et al. \cite{Raja18NSGA2DT} presented  \nsgadt, a surrogate-assisted testing technique, which is in particular a Learnable Evolutionary Algorithm (LEM) \cite{MichalskyLem2000}. This approach combines search with the evolutionary algorithm \nsga with model fitting, i.e., Decision Tree (DT) classification models. While \nsga allows to optimize fitness functions to find promising test cases, DTs support the search by dividing the search space into most likely failure-revealing or not failure-revealing regions to direct the subsequent search with \nsga into promising areas and to identify new failing tests. However, DTs impose decision boundaries that are orthogonal to the considered features, i.e., search variables, thus achieving a coarse prediction of failure-inducing regions. In particular, DTs are likely defining regions as failure-revealing even when those contain spaces without any sampled tests or non-failing tests (\autoref{fig:motivation}).

We present therefore the novel LEM-based approach \textit{\algolong} (\algo), which uses support vector machines (SVMs), a well applied ML-technique for classification tasks.  In contrast to DTs, SVMs can learn non-linear boundaries (\autoref{fig:motivation}) by transforming the original problem into a higher dimensional space using so-called kernel functions, thus allowing to separate accurately failing from non-failing test cases. 
Our presented algorithm performs the search as follows: after several iterations with \nsga, an SVM model is trained on all so far evaluated test cases to predict failure-revealing regions, followed by sampling in identified regions. Sampled test cases are then evaluated using a simulator and fitness values are assigned indicating whether the test cases are failure-revealing or not. From all identified test cases so far, the best test cases are chosen to form the seed population for the genetic search with \nsga in the subsequent iteration of \algo.

We prototypically evaluate \algo on an Automated Valet Parking system tested in the Prescan simulator~ \cite{Prescan}. Specifically, we compare the performance of \algo to \nsgadt and \rs using well-established quality indicators \cite{LiMOOEval15}. Our results show that \algo is more effective in identifying failure revealing test inputs than the compared approaches. In particular, it achieves better values for all evaluation metrics and identifies on average 34\% more distinct failures than the state-of-the-art learnable evolutionary testing approach.




\section{Preliminaries}
\label{sec:preliminaries}



\subsection{Search-based Testing Problem}
\label{def:sbst}
A search-based testing problem $P$ is defined as a tuple $P = (S, D, F, O)$, where $S$ is the system under test, $D \subseteq \mathbb{R}^n$ is the search domain, where $n$ is the $dimension$ of the search space. An element $(x_1, \ldots, x_n) \in D$ is called a test input. 
$F$ is the vector-valued fitness function defined as $F: D \mapsto \mathbb{R}^m, F(x) = (f_1(x),\ldots, f_m(x))$, where $f_i$ is a scalar fitness function (or fitness function, for short) which assigns a quantitative value to each test input, and $\mathbb{R}^m$ is the \textit{objective space}. A fitness function evaluates how \textit{fit} a test input is, assigning a \emph{fitness value} to it. $O$ is the test oracle function, $O : \mathbb{R}^m \mapsto \lbrace 0,1 \rbrace $, which evaluates whether a test passes or fails. A test that fails is called \emph{failure-revealing}.

\subsection{Support Vector Machines}
Our approach uses support vector machines (SVMs) which is a widely-applied supervised machine learning algorithm suited for classification and regression tasks \cite{07_SVM1Networks}. In the context of our algorithm, we use SVMs for classification.
An SVM aims to find a hyperplane that separates data points of different classes while maximizing the distance between the hyperplane and the nearest points of each class. SVMs can be applied both to linearly separable as well as to non-linearly separable data using the concept of a kernel function \cite{07_SVM1Networks}. In particular, a kernel function maps the original feature space into a higher-dimensional space, allowing SVMs to find a separating hyperplane in the higher-dimensional space. 

\section{SVM Guided Testing}
\label{sec:approach}

In this paper, we propose the \textit{\algolong} (\algo) a search-based testing algorithm that combines evolutionary search with SVM model learning to guide the search towards failure-revealing test cases. 

Our proposed algorithm receives as input a specification of an SBST-Problem as defined in \autoref{def:sbst}, the population size $n$, the generation size $g$ for search iterations with \nsga and a number of samples $s$ specifying how many test inputs shall be randomly created in the failure-revealing region found by the learned SVM model. In addition, the search budget is specified, which is defined by the maximal number of test input evaluations or the maximal search execution time.



\begin{figure*}[!t]
     \begin{subfigure}[b]{0.2\textwidth}
        \label{fig:cigd_example_1}
        \centering
        \includegraphics[width=0.8\textwidth]{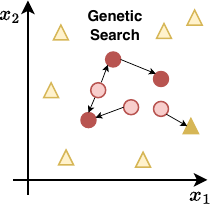}

        \caption{}
    \end{subfigure}
     \hspace{-1em}
    \begin{subfigure}[b]{0.2\textwidth}
        \label{fig:cigd_example_2}
        \centering
        \includegraphics[width=0.8\textwidth]{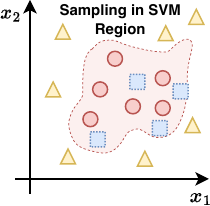}
        \caption{}
    \end{subfigure}
    \begin{subfigure}[b]{0.2\textwidth}
        \label{fig:cigd_example_2}
        \centering
        \includegraphics[width=0.8\textwidth]{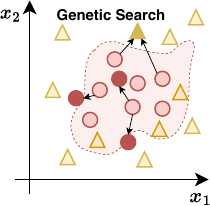}
        \caption{}
    \end{subfigure}
    \begin{subfigure}[b]{0.2\textwidth}
        \label{fig:cigd_example_2}
        \centering
        \includegraphics[width=0.8\textwidth]{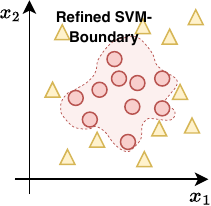}
        \caption{}
    \end{subfigure}
    \begin{subfigure}[b][0pt][b]{0.18\textwidth}
        \label{fig:cigd_example_2}
        \centering
    \raisebox{18pt}{
        \includegraphics[width=0.85\textwidth]{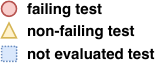}}
        \caption*{}
    \end{subfigure}
    \caption{Overview of the main steps of \algo. a) Genetic search using \nsga. b) Learned SVM-model for failing region prediction and sampling inside region. c) Genetic search with \nsga in subsequent iteration using evaluated samples and best solutions found. d) Refined SVM-model learning. Arrows indicate genetic operations.
    }
    \label{fig:algo}
\end{figure*}

\autoref{algo:svm-full} and \autoref{fig:algo} describe the main steps of \algo: Initially, test inputs are randomly generated in the whole search space $D$ forming the seed population (s. line 1-2).
An iteration of \algo begins by selecting the best, i.e., non-dominated test inputs from the set of all evaluated test cases $P$ (line 4), forming the population $Q$.
In particular, we adapt the survival operator of \nsga \cite{Deb02NSGA2}, by sorting first individuals based on whether they are failure-revealing or not, followed by ranking and finally crowding-distance sorting.
Then \nsga performs a genetic search in the domain $D$ for $g$ iterations, which yields the population $Q'$ containing all found and evaluated test cases during the search (line 5). $Q'$ is then merged with $P$ and passed to the oracle function $O$ (line 6-7).
$O$ classifies a test case as failure-revealing or non-failure-revealing, thus separating $P$ in the sets $C^+, C^-$ (line 7).


Now, by the test oracle labeled test inputs in $C^+$ and $C^-$ are passed to the SVM algorithm to train its model. Here, test inputs $x = (x_1, \ldots, x_n) \in D$ are used as features, while oracle outputs $O(x)$ are used as labels. To improve the model's accuracy, \algo employs hyper-parameter tuning using grid search and cross-validation \cite{Lameski2015gridsearch, Stone74CrossValidation, Wainer2021HowTT}, a well-applied approach for tuning hyper-parameters of an estimator. Specifically, the parameter $C$ is tuned, which controls the cost of misclassification on the training data. Further kernel-specific parameters can also be tuned.
Note, that for retraining an SVM model for different parameter combinations no repeated execution of the SUT is necessary as labels are already computed.



In the next step, new test inputs are sampled within the failure-revealing region predicted by the learned SVM model $M$ (line 9). Sampled test inputs are then evaluated and added to the set of all so far evaluated test cases $P$ (line 11). This concludes one iteration of \algo. The algorithm stops when the search budget is exhausted, and returns all failing test inputs found during search. 



\begin{algorithm}[t]
\DontPrintSemicolon
\footnotesize

    \SetKwInOut{Input}{Input}
    \SetKwInOut{Output}{Output}
    \Input{
        $(S, D, F, O)$: An SBST Problem \newline
        $n$: Population size of \nsga \newline
        $g$: Generation number of \nsga  \newline
        $s$: Number random samples to generate in SVM-region
    }
    \Output{ $C^+$: failing test cases.}
    \BlankLine
    $P \gets initialPopulation(D, n)$\tcc*{Generate initial population}
    $evaluate(P, F, S)$\tcc*{Evaluate initial population by simulation}
    \While{$search\_budget\_available$}{
        $Q \gets non\_dom\_critical(P, n)$\tcc*{Select best, failing n solutions} 
        $Q'\gets \textsc{\nsga}(Q, D, F, S, n,g )$
        \tcc*{Perform genetic search} 
        $P\gets P \cup Q'$\;
        $C^+, C^- \gets do\_label(P,O)$\tcc*{Assign labels whether failing or not.}
        $M \gets SVM(C^+,C^-)$\tcc*{Train SVM model to predict failing regions}
        $R \gets sample(M, s)$ \tcc*[r]{Sample in predicted failing region.}
        $evaluate(R,S,F)$\tcc*{Evaluate new samples R by simulation} 
        $ P  \gets P \cup R$ \;
      
    }
      $C^+, C^- \gets do\_label(P,O)$\;
    \KwRet{$C^+$}
\caption{NSGA-II-SVM}\label{algo:svm-full}
\end{algorithm}



\section{Empirical Evaluation}
\label{sec:evaluation}

For our preliminary evaluation of \algo we consider the following research question:

\textit{How effective is \algo compared to na\"ive testing and a state-of-the-art approach in identifying failure-revealing tests?}

To answer our research question we compare the performance of \algo to random search (RS) and to \nsgadt \cite{Raja18NSGA2DT}, a state-of-the-art learnable
evolutionary testing technique in the literature.
In the following we describe the SUT, the search configuration and the metrics used. 

\subsection{Experimental Setup}

\textbf{Subject System.} As case study we consider an Automated Valet Parking system (AVP)~\cite{BoschAVP}. This system allows to park a vehicle without human intervention. The AVP follows a precomputed trajectory to a free parking spot while avoiding collisions with other vehicles or pedestrians.
In particular, the SUT is a Simulink model provided by our industrial partner.
This model encompasses a Lidar sensor, a path follower, a path planner, and an automated emergency braking system. Our aim is to identify scenarios in which the AVP violates the following safety requirement provided by our industrial partner: \textit{`AVP maintains a minimum distance of $0.8m$ to other objects, such as cars or pedestrians, when the AVP vehicle's velocity is non-zero."}

To assess the AVPs conformance to this safety criteria, we employ system-level simulation-based testing in the industrial simulator Prescan~\cite{Prescan}. This exhibits the system to various scenarios~\cite{Ulbrich15scenario} to challenge the AVP.  Specifically, our focus is on a scenario where a pedestrian is crossing perpendicularly the path of the ego vehicle from the right.



\textbf{Metrics.} We adopt several quality indicators (QIs) designed to evaluate MOO algorithms as proposed by guidelines ~\cite{AliQI20, LiEvaluateSBST22, LiMOOEval15}. In particular, we evaluate QI values only on test inputs that are failure-revealing as we are only interested in those solutions.
  
\textit{Hypervolume (HV).} 
The HV indicator~\cite{Zitzler1999} measures the size of an area that is spanned by found Pareto solutions and a reference point. HV allows us to assess the convergence of an algorithm as well as the spread and uniformity of identified solutions. A larger HV value indicates a better performance by the algorithm.

\textit{Generational Distance (GD).} The GD indicator \cite{LiMOOEval15} measures the average distance between identified Pareto solutions and the real Pareto front (PF). Note, that as the real PF for our problem is not known in advance we can approximate it by union all solutions from all approaches under evaluation as proposed in the literature \cite{AliQI20}. A smaller GD value implies that the solutions set is \textit{closer} to the approximated PF.


\textit{Spread.}
The Spread (SP) indicator~\cite{Deb02NSGA2} allows us to assess how spread and uniform solutions within a PF produced by a search algorithm are. A smaller SP value indicates an evenly distributed and spread solution set.

In addition, we report on the number of on average found \textit{distinct} failing tests after the last evaluation. To mitigate the bias of having \textit{close} solutions, we adopt the approach from Zohdinasab et al. \cite{Zohdinasab21DeepHyperion} and divide the oracle-constrained area in the objective space into a grid of cells of equal size. Then we assign each solution based on its fitness values to a cell in the grid. Two solutions are called \textit{distinct} if and only if they are located in different cells. For the evaluation, we divide each dimension into 50 cells.


\textbf{Search Configuration.} The search domain for the AVP Case Study is defined by three search variables: $x_1$, the velocity of the ego vehicle, $x_2$, the velocity of the pedestrian, and $x_3$, the time at which the pedestrian starts moving. The range $D_i$ for each search variable is defined as follows: $D_1 = [0.1 m/s, 3 m/s]$, $D_2 = [0.5 m/s,2 m/s]$ and $D_3 = [0s,5s]$.

\textit{Fitness functions.} We use two fitness functions: $f_1$ measures the adapted distance between the ego vehicle and the pedestrian\footnote{Note, that $f_1$ combines the longitudinal and latitudinal distance between the front of the ego vehicle and another actor in one value. A value of 1 is reached when both actors collide, while it approaches 0 with increasing distance.}, while $f_2$ is the velocity of the ego vehicle at the time of the minimal distance between the ego vehicle and the pedestrian.
As oracle function $O$, we use the constraint $(f_1 > 0.8) \wedge (f_2 > 0.1)$, which specifies that ego violates a safety distance, moving at the same time with a \textit{small} velocity of minimal 0.1 m/s,

\textit{Algorithm Configuration.} For \algo and \nsgadt we set the population size to 20. We choose a small population size to allow the algorithm to perform many iterations of constructing ML/DT models to be able to recognize the effect of using an ML-technique.
Further, we use equal crossover and mutation parameter values. As selection operator, we use binary tournament selection as originally proposed in \nsga. We set the number of \nsga generations for \nsgadt to 5 and set DT parameters such as the impurity ratio and \textit{criticality} threshold as proposed by Abdessalam et al. \cite{Raja18NSGA2DT}. For \algo, we set the number of random samples after each SVM iteration to 30 and use rejection sampling~\cite{luke2012essentials}. That means, to generate new samples we sample in the whole search space until the required number of samples is found within imposed boundaries.

As kernel function in \algo we have selected the radial basis kernel (RBF) \cite{buhmann2003rbf}, as preliminary evaluations using an RBF kernel have shown promising results compared to other well-applied kernels (i.e., sigmoid or polynomial kernel \cite{Patle2013kernel}). For the parameter optimization of $\gamma$ and $\mathcal{C}$ of the SVM model, we perform 5-fold cross-validation and use logarithmically spaced values $[1,10,100,1000]$ and $[0.01,0.1,1,10]$ for the grid search, as proposed by several works \cite{RSandGridOptBergstra, syarif2016svm, Wainer2021HowTT}. To generate a diverse initial population we use for both search approaches Latin Hypercube Sampling \cite{LHS}, a  well-applied quasi-random sampling technique. We let both algorithms run for a minimum of 1,000 evaluations as preliminary experiments have shown that metric values already stabilize before. For a comparison with RS we generate 1,000 test inputs. 
For all algorithms, we repeat the experiments 10 times to account for their randomness.

\vspace*{-7pt}

\subsection{Results}
\begin{figure}
    \centering
    \includegraphics[scale=0.5]{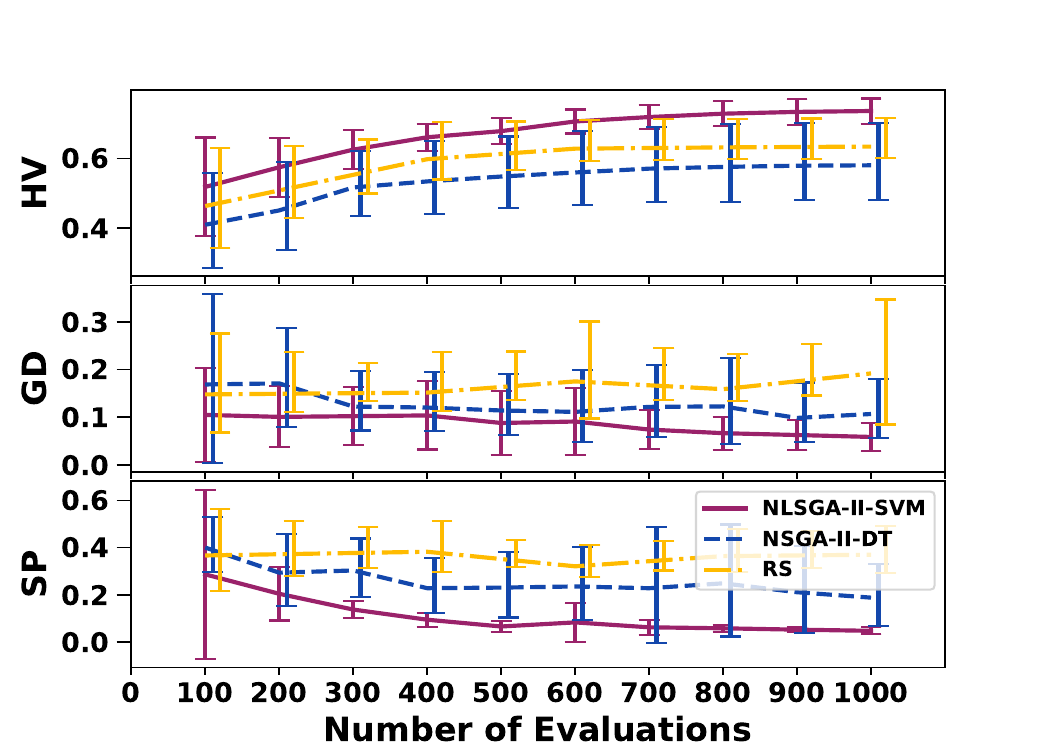}
    \caption{ Quality metric results averaged over 10 runs.}
    \label{fig:results-qi}
\end{figure}

 The resulting quality indicator values are shown in \autoref{fig:results-qi}. We report the averaged scores of the evaluation metrics and the standard deviations after every 100 evaluations.
 
\textbf{Comparison QI results.}
Compared to RS, our proposed algorithm achieves throughout the search better values for all QIs. Also, the variation in the QI values is for \algo significantly lower and decreases with an increasing number of evaluations.

Regarding the comparison of \algo to \nsgadt, we can see that for all QIs \algo has always a smaller variation then \nsgadt. The variations for GD and Spread decrease for \algo at a high rate, while for \nsgadt the deviations are almost equal over time. Since the QI values for \nsgadt seem to stabilize at a certain value, the QI values are unlikely to improve in case we would have run \nsgadt longer.

Regarding the number of failure-revealing tests found, over all runs, we identify on average for \algo 93 distinct failing tests, while for \nsgadt the number is 69 and for RS 28.


\textbf{Statistical tests.} To evaluate the statistical significance of the results we perform statistical testing using the pairwise Wilcoxon Paired Signed Ranks test and Vargha and Delaney’s $\hat{A}_{12}$ effect size measure~\cite{Arcuri14statSE}. The significance level has been set to 0.05, following existing guidelines \cite{Arcuri14statSE}. We chose QI values after 1,000 evaluations for the comparison. The comparison yields for all QIs p-values below 0.05 with a large effect size for HV \cite{replication-package}. For GD we get a small and for Spread a negligible effect size. Overall, the results indicate that differences for the selected QI values of \algo and baselines are statistically significant.


\begin{tcolorbox}
Our answer to the research question is that \algo is more effective in identifying failing test inputs than na\"ive \rs as well as a state-of-the-art learnable evolutionary testing technique as it identifies on average 34\% more distinct tests and yields better QI values when applied to our preliminary industrial ADS Case Study.
\end{tcolorbox}

\section{Threats to Validity}
\label{sec:usage}

To mitigate \textit{internal validity} risks, we used for all experiments with baseline approaches the same search space and the same SUT. The SUT is a real-world ADS. The ADS and the definition of its search space are provided by an industrial partner. Furthermore, we have used a high-fidelity simulator Prescan to simulate the ADS.




Regarding \textit{construct validity}
we used well-applied quality indicators HV, GD and SP from the search-based software engineering domain to assess the effectiveness of our search approach. These metrics reflect the effectiveness of SBST algorithms as shown by Li et al. \cite{LiEvaluateSBST22}. However, GD metric results depend on the quality of the approximation of the real PF. In our evaluation, the approximation was done by the union of all solutions from all runs which might be less representative compared to a PF gained by more runs.



Regarding \textit{external validity}, we tested an AVP with a relatively \textit{small} search space (three search variables, small ranges). It requires experiments with more complex systems from different domains. Also, the SVM model's accuracy depends on the size of the training set. The guidance of the model on a system that fails only on a small number of inputs might be not as effective as in our study.

To allow \textit{replicability}, we implemented \algo using the open-source framework OpenSBT \cite{sorokin2023opensbt} and made results of this paper publicly available \cite{replication-package}.

\section{Related Work}
\label{sec:related-work}

ML-driven testing has received much attention in the recent past. Several approaches \cite{Raja16NeuralNSGA2,Raja18NSGA2DT, Haq2022SurrogateMOP,zhong2023adfuzz} have been proposed employing ML models to avoid costly evaluations of test cases by approximating fitness values or by guiding the search process.
Haq et al. \cite{Haq2022SurrogateMOP} for instance combine so-called local and global surrogate models to replace expensive simulations. Abdessalam et al. \cite{Raja16NeuralNSGA2} use neural networks with MOO to reduce the search time bypassing costly test case evaluations by training neural networks for fitness value prediction. However, our approach is not focused on replacing simulation executions but on guidance towards promising inputs that should be considered for evaluation. 



AutoFuzz~\cite{zhong2023adfuzz} is focused on fuzzing the test scenario specification by using a seed selection mechanism based on a neural network classifier that selects likely traffic-violating seed inputs as well as an ML-based adaption of the mutation. 

\nsgadt \cite{Raja18NSGA2DT} is related to our approach but tailors DTs instead SVMs to direct the search toward promising tests. 
Besides the used classification model, a further significant difference between \algo and \nsgadt is that in \nsgadt the guidance by DT is enabled by initiating \nsga search using failing tests from identified failure-revealing leafs. In contrast, in \algo we sample inside the region predicted by the SVM model to define the initial population for \nsga.
This implies that in \nsgadt, the execution time is not controllable because the population size in \nsga is only bounded by the number of in total produced tests, i.e., number of failing tests found in a leaf. In contrast, our approach maintains a constant population size.

\section{Discussion and Future Work}
\label{sec:conclusion}

\balance
We have presented a surrogate-assisted testing approach that combines evolutionary search with support vector machine classification models to guide the search towards failing test inputs.
A preliminary evaluation of our approach on an automated driving system has shown that our approach outperforms the selected baseline techniques. However, in our evaluation, we employ rejection sampling to explore test inputs in the predicted region by the SVM model. Even though evaluating every single sample in the simulator is not required, sampling in a high dimensional search space with relatively small failure-revealing regions can be time-consuming \cite{luke2012essentials}. In addition, the execution time of \algo depends on the time for the parameter tuning using grid search. As shown by Lameski et al. \cite{Lameski2015gridsearch}, heuristic search approaches are viable alternatives for finding parameters for more accurate SVM models.

In our future work, we want to improve the sampling strategy within identified failure-revealing regions and evaluate the approach on various deep learning-enabled systems from different domains with more complex search spaces.

\section*{Acknowledgments}
\includegraphics[scale=0.018]{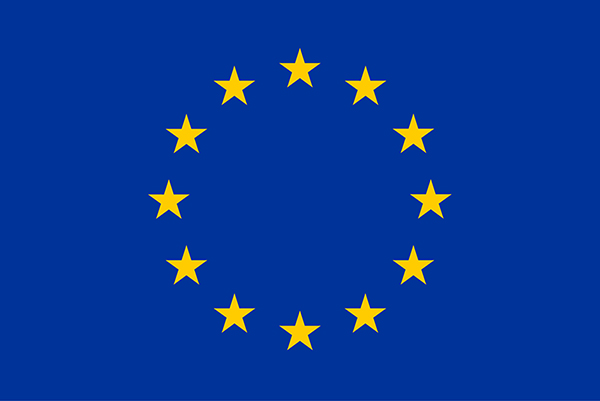}  This paper has received funding from the European Union’s Horizon 2020 research and innovation programme under grant agreement No 956123, and the NSERC of Canada under the discovery grant. 
\printbibliography

\end{document}